\documentclass[12pt,reqno]{article}

\usepackage{amsmath, amsthm, amssymb,epsfig,alltt}

\def\a{\alpha}
\def\b{\beta}
\def\bt{{\beta_T}}

\def\e{\epsilon}
\def\p{\partial}

\def\t{\tau}
\def\th{\theta}
\def\s{\sigma}

\def\o{\omega}

\def\half{\frac{1}{2}}

\def\nn{\nonumber}

\def\2pap{2\pi\alpha^\prime}

\def\beq{\begin{eqnarray}}
 \def\eeq{\end{eqnarray}}
 \def\4pap{4\pi\a^\prime}

 \def\ap{{\a^\prime}}

 \def\ta{{\tilde \a}}
 \def\tb{{\tilde \b}}

 \def\bole{{\boldsymbol \epsilon}}
 \def\bolth{{\boldsymbol \theta}}

\begin{document}


\title{Quons in a Quantum Dissipative System}

\author{Taejin Lee \\~~\\
Department of Physics, Kangwon National University, \\
Chuncheon 200-701 Korea and~\\
email: taejin@kangwon.ac.kr}

\maketitle

\centerline{\bf Astract}

String theory proves to be an imperative tool to explore the critical behavior of the quantum dissipative system. We discuss the quantum particles moving in two dimensions, in the presence of a uniform magnetic field, subject to a periodic potential and a dissipative force, which are described by the dissipative Wannier-Azbel-Hofstadter (DWAH) model.
Using string theory formulation of the model, we find that 
the elementary excitations of the system at the generic points of the off-critical regions, in the zero temperature limit are quons, which satisfy q-deformed statistics. 



\vskip 2cm

\section{INTRODUCTION}

The quantum mechanical description of dissipation has been one of 
the outstanding problems in theoretical physics. Since the well-known dissipative or
frictional force term in the classical equation of motion cannot be driven 
from a local action, it had been a conundrum how to quantize the dissipative 
system for a long time until Caldeira and Leggett \cite{caldeira83ann,caldeira83phy}
provided a proper answer to this problem by coupling a bath or environment which consists of an infinite number of harmonic oscillators to the system. 
In the quantum theory, the interaction with the bath produces a non-local effective
interaction, and the dissipative quantum system exhibits a phase transition, unlike one-dimensional quantum mechanical systems with local interactions only. 

If a uniform mangetic field is introduced to the system, the phase diagrams of 
the quantum dissipative system become more complex and interesting. The model, describing quantum particles subject to a uniform magnetic field, a periodic potential and a dissipative force is called the dissipative Wannier-Azbel-Hofstadter model (DWAHM)
\cite{dwahm}.
The DWAHM has been studied extensively for more than two decades \cite{callan91,Callan:1992vy,freed93,Callan:1995ck}, 
since it is known to have a wide range of 
applications which include
Josephson junction arrays \cite{larkin,fazio,sodano},
the Kondo problem \cite{Affleck:1990by}, the study of 
one-dimensional conductors \cite{Kane:1992a}, 
and junctions of quantum wires~\cite{Oshikawa:2005fh}. 


The non-local dissiative term of DWAHM can be traded with the local Polyakov term if
the time dimension of the model is mapped onto the boundary of the disk diagram in string theory. The bulk degrees of freedom of the string play the role of the bath to 
produce the non-local dissipative term at the boundary. 
The advantage of this approach is clear in that we only have to deal with a local two-dimensional quantum field theory where string theory techniques are readily available. Along this line the model was first discussed by
Callan and Thorlacius \cite{callan90}. The fractional statistics, or the interchange phase arising from the interaction with magnetic field, has been analyzed in the framework of the Coulomb gas expansion of 
the boson theory in ref.\cite{Callan:1995ck}. 
Here, we will further extend the work in this direction, to study the critical behavior of the DWAHM and its elementary excitations in worldsheet fermion theory,
applying the recent developments in string theory.

\section{The Qauntum Dissipative System:\\
The Schmid Model}

In the absence of the magnetic field, the DWAHM reduces to the Schmid model \cite{schmid} which has been adopted to study the quantum Brownian motion \cite{fisher;1985} and 
the transport of a one-channel Luttinger liquid \cite{kane;1992b,furusaki}. The action for a particle, subject to a dissipative force is given as follows:
\beq 
S &=& \frac{\eta}{4\pi \hbar} \int^{\bt/2}_{-\bt/2} dt dt^\prime 
\frac{\left(X(t) - X(t^\prime)\right)^2}{(t-t^\prime)^2} 
+ \frac{M}{2\hbar} \int^{\bt/2}_{-\bt/2} dt \dot X^2
+ \frac{V_0}{\hbar} \int^{\bt/2}_{-\bt/2} dt \cos \frac{2\pi X}{a}. 
\eeq
The first term is responsible for the dissipation, and the second 
term is the usual kinetic term for a particle with mass $M$. The 
third term denotes a periodic potential. 
If we wish to study the system in real time, we
may take the limit where $\bt=1/T \rightarrow \infty$. Since we are only 
interested in the long-time behavior of the system, we may ignore
the kinetic term, which only plays a role of regulator in the 
long-time analysis. 

Mapping the Schmid model to the string theory on a disk
begins with identifying the time $t$ as the boundary parameter $\sigma$ 
in string theory.
The action of the Schmid model precisely coincides with the boundary 
effective action for the bosonic string subject to a boundary periodic potential
on a disk with a boundary condition $X(\t=0,\s) = X(\s)$, $\s= 2\pi t/\bt$, 
\beq
S =  \frac{1}{4\pi \ap}
\int d\t d\s \p_\a X \p^\a X - \frac{V}{2}
\int d\s \left(e^{iX}+ e^{-iX} \right). 
\eeq 
The physical parameters of the two theories are identified as 
$\frac{1}{4\pi \ap}=\frac{\eta}{2 \hbar} \left(\frac{a}{2\pi}\right)^2,\, 
V= \frac{V_0}{\hbar} \frac{\bt}{2\pi}$.
The Schmid model has two phases. In the region where $\ap > 1$, the boundary periodic interaction becomes an irrelevant operator. Thus, it can be ignored in the low energy physics, which we are mostly interested in, and the quantum particle can be described by the free bosonic string action. In this phase, 
the quantum particle is delocalized. On the other hand, 
in the region where $\ap <1$, the boundary interaction, becoming a relevant operator, 
dominates in the action. In this phase, the quantum particle is localized and the physical degrees of freedom are kinks, which depict the tunneling of the quantum particle between adjacent minima of the periodic potential. The model possesses an interesting duality between particles and kinks; the theory with
$\ap$ in the strong periodic potential regime is mapped to its dual theory with $1/\ap$ in the weak periodic potential regime. 

At the point where $\ap = 1$, {\it i.e.}, the boundary periodic potential becomes marginal. 
The most efficient way to calculate the partition function and the correlation functions of 
physical operators is to make use of the boundary state formulation. The boundary state formulation of the 
Schmid model at the critical point has been studied in refs.\cite{Callan:1994ub,Polchinski:1994my}.
Later it has been studied as a conformal field theory for the rolling tachyon in string theory \cite{Sen:2002nu,Tlee:06,TLee;08}, in connection with the decay of unstable objects called D-branes.  The partition function for the quantum particle at the critical point may be obtained by
\beq
Z &=& \int D[X] e^{-S_{SM}[X]} = \langle 0| B\rangle, \quad
|B \rangle = \int D[X] \exp \left[\frac{V}{2} \int d\s \left(e^{iX} + e^{-iX} \right)
\right]|X \rangle 
\eeq
where $|X\rangle$ is the eigenstate of the string operator $\hat{X}$, 
$\hat{X}|X\rangle = X|X\rangle$ and $|0\rangle$ is the ground state of string theory.
This partition function can be exactly calculated if we rewrite the action in terms of fermion fields using the Fermi-Bose equivalence.
The periodic interaction term corresponds to a bilinear fermion operator 
\cite{Lee:2005ge}
\beq
e^{iX} + e^{-iX} =  \sum_{i=1}^2 \left(\psi^\dagger_{iL} \psi_{iR} + \psi^\dagger_{iR} \psi_{iL}\right)
\eeq
where
\beq \label{fermion}
\psi_{iL} &=& \eta_{iL} :e^{-i\sqrt{2} \phi_{iL}}, \quad
\psi_{iR} = \eta_{iR} :e^{i\sqrt{2} \phi_{iR}}, \nn\\
\phi_{1 L/R} &=& \frac{1}{\sqrt{2}}(X_{L/R}+Y_{L/R}),\quad \phi_{2L/R} = \frac{1}{\sqrt{2}}(X_{L/R}-Y_{L/R}),
\eeq
and $\eta_{iL/R}$ are cocycles. (Here, an auxiliary free boson field $Y$ is introduced. 
We choose the Dirichlet condition as the boundary condition for $Y$ so that it is 
decoupled from the physical degrees of freedom $X$.) 
This observation indicates that at the critical point the elementary excitations on the boundary are fermions. 
Off the critical point, the bulk action can be rewritten as an action for interacting fermions, {\it i.e.} the 
Thirring model action. Thus, the elementary excitations of the Schmid model can be understood as fermions.   

\section{The Dissipative Wannier-Azbel-Hofstadter model }

If we turn on the magnetic field, the Schmid model becomes the DWAHM, of which action
is given by
\beq 
S &=& \frac{\eta}{4\pi \hbar} \int^{\bt/2}_{-\bt/2} dt dt^\prime 
\frac{\left(X^I(t) - X^I(t^\prime)\right)^2}{(t-t^\prime)^2} + \frac{ieB_H}{2\hbar c} \int^{\bt/2}_{-\bt/2} dt \epsilon^{IJ}
\p_t X^I X^J \nn\\
&& + \frac{V_0}{\hbar} \int^{\bt/2}_{-\bt/2} dt \sum_I
\cos \frac{2\pi X^I}{a},
\eeq
where $I,J = 1, 2$. 
The action for the DWAHM can be interpreted 
as the boundary effective action 
for the closed string on a disk with a periodic tachyon potential and 
the Neveu-Schwarz (NS) B field with a boundary condition $X^I(\t=0,\s) = X^I(\s)$ \cite{callan91}.
\beq
S&=&  \frac{1}{4\pi}
\int d\t d\s E_{IJ} \left(\p_\t + \p_\s\right) X^I 
\left(\p_\t - \p_\s\right) X^J - \frac{V}{2} \int d\s \sum_I \left(e^{iX^I}+ 
e^{-iX^I}\right)
\eeq 
where $E_{IJ} = \a\delta_{IJ} + 2\pi B_{IJ}= \a\delta_{IJ} + \b \epsilon_{IJ}$, 
$\a = 1/\ap$ and $2\pi\b = \frac{eB_H}{\hbar c}a^2$.
The boundary state corresponding to the DWAHM may be written as 
\beq 
|B\rangle &=& \,\exp\Biggl[\frac{V}{2} \int_{\p M} d\s\sum_I \left(
e^{iX^I} + e^{-iX^I} \right) \Biggr]|B_E\rangle . \label{boundary}
\eeq 
The boundary state $|B_E\rangle$ satisfies the boundary condition
\beq
\left(E_{IJ} \a^J_{-n} + E^T_{IJ} \ta^J_n\right) |B_E\rangle= 0, 
\quad p^I |B_E\rangle= 0
\eeq
as given in terms of the oscillators of $X^I$
\beq 
X^I(\s,0) &=& x^I + \o^I \s  +i\sqrt{\frac{1}{2}} \sum_{n\not=0} \frac{1}{n}
\left[\a^I_n e^{in\s} + \ta^I_n e^{-in\s}\right]
\eeq
where the normal mode operators satisfy the canonical commutation relations as
\beq
\left[ x^I,p^J\right]&=&i\delta^{IJ},~
\left[ \a^I_m, \a^J_n \right] = g^{IJ}m\delta_{{m+n},0},\nn\\ 
\left[\tilde\a^I_m, \tilde\a^J_n \right] &=& g^{IJ} m\delta_{{m+n},0}, \quad g^{IJ} = \a^{-1}\delta^{IJ}.
\eeq
Here $\o^I$, $I=1,2$, are winding numbers.

The boundary state $|B_E \rangle$ has a simpler expression if we choose a new 
oscillator basis $\{\beta^I_n, \bar\beta^I_n\}$ which is related to the basis 
$\{\a^I_n, \bar\a^I_n\}$ by the $O(2,2,R)$ transformation  generated by $T$ \cite{Lee;mod,Lee;canonical,Lee;2007}
\beq 
T &=& \left(\begin{array}{cc} I & 0 \\ \bolth/(2\pi) & I \end{array} \right),\label{t1}\\
\bolth/(2\pi) &=& \frac{1}{E} (2\pi B) \frac{1}{E^T} = \frac{\b}{\a^2+\b^2} \bole,\label{t2}\\
\a^I_n &=& g^{IJ} E_{JK} \b^K_n,\quad \ta^I_n = g^{IJ} E^T_{JK} \tb^K_n,\label{t3}
\eeq

In the new oscillator basis, the boundary condition 
is transcribed into the Neumann condition as
\beq \label{neumann}
\left(\b^I_{-n} + \tilde \b^I_{n}\right) |B_E\rangle = 0.
\eeq
Note that the oscillators $\{\beta^I_n, \bar\beta^I_n\}$ 
respect the worldsheet metric $G$
\beq
G_{IJ} &=& (E^T g^{-1} E)_{IJ}= \left(\frac{\a^2+\b^2}{\a}\right) \delta_{IJ}, \label{metric}\\
\left[\b^I_n, \b^J_m\right] &=& (G^{-1})^{IJ} n  \delta(n+m), ~~~
[\tb^I_n, \tb^J_m] =(G^{-1})^{IJ} n \delta(n+m)
\eeq
and the string coordinate operators $X^I$
are no longer commuting operators in the new basis
\beq
\left[X^I(\s_1), X^J (\s_2) \right] =  -\frac{2\b}{\a^2+\b^2} \e^{IJ} \sum_{n\not =0} \frac{1}{n} e^{in(\s_1-\s_2)} = \frac{2i}{\pi} \th^{IJ} \arctan
\left[ \frac{\sin(\s_1-\s_2)}{\cos(\s_1-\s_2) -1} \right].
\eeq
In the zero temperature limit, where $\bt \rightarrow \infty$,
\beq \label{noncomm}
x \rightarrow -2/(\s_1-\s_2) = -\bt/\pi(t_1-t_2). 
\eeq
From $\tan(\pi/2 \pm 0) = -\infty$, it follows that in the zero temperature limit
\beq
\left[X^I(\s_1), X^J (\s_2) \right] = i \th^{IJ}.
\eeq
This is precisely the non-commutative relation between the open string coordinate operators 
\cite{seib,Lee;open,lee0105}.
In the open string theory the algebra of the coordinate operators, defined 
at equal $\t$ at end points is non-commutative. In closed string theory, as the world sheet parameters
are interchanged, these points are on the boundary $\t=0$ at equal $\s$. Thus, the non-commutative
algebra of open string is expected to emerge in the low temperature limit or the equal $\s$ limit in the 
closed string theory. 
Since $\tan$ function has a periodicity of $\pi$, we may also make an alternative choice;
$\tan (\pi/2 \pm 0) = \mp\infty$. It yields in the zero temperature limit
\beq \label{alternative}
\left[X^I(\s_1), X^J (\s_2) \right] = i \th^{IJ}\, {\rm sign}(\s_1-\s_2). 
\eeq
However, this choice does not agree with the non-commutative algebra obtained in the open string picture.
It also results in an inconsistent operator algebra as we will see shortly. 
 
When we expand the boundary state or the partition 
function as powers of $V$,
this non-commutativity of the operators $X^I$ produces in the integrands, 
powers of the phase factor $e^{i\th}$ which reduce to $1$ if $\frac{\th}{2\pi}$ is an integer.
From this it follows that on the circles of the two dimensional plane of
$(\a,\b)$ where
\beq
\a^2 + \left(\b - \frac{1}{2n}\right)^2 = \left(\frac{1}{2n}\right)^2, ~~~ n \in Z
\eeq
the magnetic field can be completely removed by the $O(2,2;R)$ transformation generated by $T$, 
Eq.(\ref{t1}) 
and the DWAHM can be mapped into the string theory with the world sheet metric $G_{IJ}$, Eq.(\ref{metric})
on a disk with the periodic tachyon potential only 
({\it i.e.}, the Schmid model) \cite{Lee;mod,Lee;2007}. These circles are termed as magic circles \cite{callan91}.

At the generic points of the phase space $(\a,\b)$, the coordinat operators $X^I(\s,0)$ at the 
boundary may be written as 
\beq
X^I(\s,0) &=& Z^I(\s,0) + \frac{i}{\sqrt{2}} \frac{\b}{\a} \sum_{n\not=0} \frac{1}{n}
\e^{IJ} \left(\b^J_n+\tb^J_{-n} \right) e^{in\s}, \\
Z^I(\s,0) &=& x^I + \o^I \s + i\frac{1}{\sqrt{2}} \sum_{n\not=0} \frac{1}{n}
\left[\b^I_n e^{in\s} + \tb^I_n e^{-in\s}\right].
\eeq
Here $Z^I$, $I=1,2$, are commuting coordinate operators of the closed string with the world sheet metric 
$G_{IJ}$. In order to evaluate the boundary state and the partition function succintly it is necessary to 
rewrite the periodic potential as a bilnear operators as in the case of the Schmid model. The periodic potential term can be written as a bilinear operators of the following operators, which become usual 
fermi field operators in the absence of the magnetic field, 
\beq
\Psi^I_{1L} &=& \eta^I_{1L} :e^{-i(X^I_L+Y^I_L)}:, \quad 
\Psi^I_{2L} = \eta^I_{2L} :e^{-i(X^I_L-Y^I_L)}: \nn\\
\Psi^I_{1R} &=& \eta^I_{1R} :e^{i(X^I_R+Y^I_R)}:, \quad 
\Psi^I_{2R} = \eta^I_{2R} :e^{i(X^I_R-Y^I_R)}:.
\eeq
Here $\eta^I_{iL/R}$ ($I=1,2$ $i=1,2$) are cocycles \cite{TLee;08}, which ensure the usual fermion anti-commutation relations between the operators $\Psi^I_{iL/R}$ when the magnetic field is turned off and
\beq
X^I_L = Z^I_L + \frac{i}{\sqrt{2}} \frac{\b}{\a} \sum_{n\not=0} \frac{1}{n}
\e^{IJ} \b^J_n e^{in\s}, \quad
X^I_R = Z^I_R -\frac{i}{\sqrt{2}} \frac{\b}{\a} \sum_{n\not=0} \frac{1}{n}
\e^{IJ} \tb^J_{n} e^{-in\s}.
\eeq
Following the previous cases of the Schmid model and the rolling tachyon model, 
we introduce auxiliary boson fields $Y^I$.
The bulk action for $Y^I$ has the Ployakov term only with the world sheet metric $G_{IJ}$. 
The Dirichlet condition is chosen as
the boundary conditions for $Y^I$, so that $Y^I$ would be decoupled from the 
physical degrees of freedom.  

Using the identity, $e^A e^B = e^B e^A e^{[A,B]}$ and the non-commutative relations for the chiral 
bosons, in the zero temperature limit
\beq
\left[X^I_{L/R}(\s_1), X^J_{L/R}(\s_2)\right] = i\th^{IJ}/2,
\eeq
we have 
\beq
\Psi^I_{iL/R}(\s) \Psi^J_{jL/R}(\s^\prime) &=& - e^{- \frac{i}{2} \th^{IJ}} \Psi^J_{jL/R} 
(\s^\prime) \Psi^I_{iL/R} (\s) \nn\\
\Psi^{I\dagger}_{iL/R}(\s) \Psi^J_{jL/R}(\s^\prime) &=& -e^{\frac{i}{2} \th^{IJ}} \Psi^J_{jL/R} (\s^\prime) \Psi^{I\dagger}_{iL/R} (\s) + 2\pi \delta^{IJ}\delta_{ij} \delta(\s-\s^\prime) \label{exotic}\\
\Psi^{I\dagger}_{iL/R}(\s) \Psi^{J\dagger}_{jL/R}(\s^\prime) &=& - e^{- \frac{i}{2} \th^{IJ}} 
\Psi^{J\dagger}_{jL/R} (\s^\prime) \Psi^{I\dagger}_{iL/R} (\s) \nn
\eeq
and 
$\{\Psi^I_{iL}, \Psi^J_{jR}\} = \{\Psi^{I\dagger}_{iL}, \Psi^J_{jR}\}
=\{\Psi^I_{iL}, \Psi^{J\dagger}_{jR}\}=0$. 
The field operators $\Psi^I_{iL/R}$ satisfy the q-deformed statistics in the zero temperature limit.
Here we should note that the q-deformed algebra of the field 
operators $\Psi^I_{iL/R}$ would be inconsistent if the alternative commutators for $X^I$, Eq.(\ref{alternative}) are chosen instead of Eq.(\ref{noncomm}). 
On the magic circles the algebra Eq.(\ref{exotic}) reduces to the commuting one if $\th/2\pi$ is an odd integer or the anti-commuting one 
if $\th/2\pi$ is an even integer
\beq
[\Psi^I_{iL}(\s_1), \Psi^J_{jL}(\s_2)]_\pm &=
& [\Psi^{I\dagger}_{iL}(\s_1), \Psi^{J\dagger}_{jL}(\s_2)]_\pm = 0\nn\\
~
[\Psi^{I\dagger}_{iL}(\s_1), \Psi^J_{jL}(\s_2)]_\pm &=& 2\pi \delta^{IJ} 
\delta_{ij} \delta(\s_1-\s_2).
\eeq
Hence, on the magic circles, $\Psi^I$ behave as para-Fermi fields \cite{para} or Fermi fields.
At generic points the operators $\Psi^I$, which describe the excitations at the boundary, satisfy the q-deformed statistics. 
Depending on $\th$, the statistics of $\Psi^I$ interpolates between the Fermi-Dirac and the para-Fermi statistics.

If we rewrite the bulk action in terms of $Z^I$, it takes the form of free string theory on a two-dimensional space with the metric $G_{IJ}$. Since the boundary state $\vert B\rangle$ is expanded in powers of $V$ as
follows
\beq
|B\rangle = \sum_n \left(\frac{V}{2}\right)^n \left(\int_{\p M} d\s\sum_I \left(
e^{iX^I} + e^{-iX^I} \right)\right)^n|B_E\rangle ,
\eeq
up to the first order of $V$, it can be written as
\beq
|B\rangle &=& |B_E\rangle + \frac{V}{2} \int_{\p M} d\s\sum_I \left(
e^{iX^I} + e^{-iX^I} \right)|B_E\rangle + {\cal O}(V^2),\nn\\
&=& |B_E\rangle + \frac{V}{2} \int_{\p M} d\s\sum_I \left(
e^{iZ^I} + e^{-iZ^I} \right)|B_E\rangle + {\cal O}(V^2).
\eeq
Thus, in the first order of $V$, the DWAHM can be mapped onto the closed string theory with the metric
$G_{IJ}$ and the periodic potential, {\it i.e.}, the Schmid model, which is equivalent to the Thirring model
with a boundary mass. This observation leads us to the the renormalilzation group flow of $V$, 
following from the perturbative analysis of the Thirring model \cite{TLee;08};
\beq V= V_0 \left(\Lambda^2/\mu^2\right)^{\frac{\a^2+\b^2-\a}{2(\a^2+\b^2)}}.
\eeq

As in the case of the Schmid model the DWAHM has two phases. In the inner region of the critical circle
\beq
\left(\a-\half\right)^2 +\b^2 = \left(\half\right)^2, \label{circle}
\eeq
the boundary periodic potential becomes an irrelevant operator and the quantum particle is delocalized. 
In the outer region of the critical circle the periodic potential, 
become relevant and the quantum particles are localized at the minima of the potential. In the localized 
phase, the solitons, called kinks become dynamical objects and new degrees of freedom $\widehat{X}^I$, dual 
to $X^I$ emerge. It is a generalized particle-kink duality of the Schmid model. The dual coordinate fields
$\widehat{X}^I$ respect the metric $\hat{G}_{IJ}=(G^{-1})_{IJ}$ and the non-commutativity 
parameter $\hat\th = 2\pi \b$ \cite{Lee;mod}. 
Thus, at generic points on the phase space, the elementary excitations
of the model in the localized phase are also quons.

\section{Conclusions}

The various forms of q-deformed statistics \cite{quon} have been proposed and studied in numerous literature. However, it is very rare to find their concrete 
realizations in realistic models. In this letter, we show that the quons, which satisfy the generalized q-deformed statistics, may be realized as elementary excitations of the realistic quantum dissipative system at 
generic points of phase space. 
The intimate relation between the quantum dissipative system and the 
string theory enables us to explore the critical behavior of the quantum dissipative 
system described by the DWAHM. By the $O(2,2;R)$ target space duality transformation of string theory,
the magnetic field term is removed from the bulk action. At the same time algebra of the string coordinate operators $X^I$ become non-commutative in the zero temperature limit or in the equal $\s$ limit. 
It is the closed string theory realization fot the non-commutativity which is mainly discussed in the 
context of the open string theory. The field operators $\Psi^I_{L/R}$, of which the periodic potential
at the boundary is written as bilinear products, then satisfy the q-deformed algebra. 

It would be most 
efficient to evaluate the partition function and correlation functions in terms of the quon fields
operators $\Psi^I_{L/R}$. Immediate applications of this work may be found in condensed matter physics,
since the DWHAM action is known to have a wide range of applications, such as Josephson junction arrays, the Kondo problem and junctions of quantum wires. Therefore, the quons are also anticipated to play 
important roles in those interesting condensed matter systems. Since the DWAHM model is equivalent 
to the rolling tachyon model in the presence of the NS B-field, we can apply this work also to the string 
theory to study the decay of unstable D-branes in the NS B-field background.

\vskip 1cm

\noindent{\bf Acknowledgments}
This work was supported by Kangwon National University.


%

%








\begin{thebibliography}{0}

\bibitem{caldeira83ann} 
A. O. Caldeira and A. J. Leggett, 
Ann. Phys. {\bf 149}, 374 (1983). 

\bibitem{caldeira83phy}
A. O. Caldeira and A. J. Leggett,
Physica {\bf 121A}, 587 (1983).

\bibitem{dwahm}
M. Azbel, Zh. Eksp. Teor. Fiz. {\bf 46}, 929 (1964) [Sov. Phys. JETP {\bf 19}, 634 (1964)]; D. R. Hofstadter, Phys. Rev. B {\bf 14}, 2239 (1976); G. H. Wannier, 
G. M. Obermair and R. Ray. Phys. Status Solidi (b) {\bf 93}, 337 (1979).

\bibitem{callan91}
C. G. Callan, Jr. and D. Freed,
Nucl. Phys. B {\bf 374}, 543 (1992).

\bibitem{Callan:1992vy}
C.~G.~Callan, A.~G.~Felce and D.~E.~Freed,
Nucl.\ Phys.\ B {\bf 392}, 551 (1993).

\bibitem{freed93}
D. E. Freed,
Nucl. Phys. B {\bf 409}, 565 (1993).

\bibitem{Callan:1995ck}
C.~G.~Callan, I. R. Klebanov, J. M. Maldacena and A. Yegulalp,
Nucl. Phys. B {\bf 443}, 444 (1995).



\bibitem{larkin}
L.I.~Glazman and A. I.~Larkin, 
Phys. Rev. Lett. {\bf 79}, 3736 (1997).

\bibitem{fazio}
R. Fazio and H. van der Zant, 
Phys. Rep. {\bf 355}, 235 (2001).
 
\bibitem{sodano}
D.~Giuliano and P.~Sodano, 
Nucl. Phys. B {\bf 711}, 480, (2005).


\bibitem{Affleck:1990by}
I.~Affleck and A.~W.~W.~Ludwig,
Nucl.\ Phys.\ B {\bf 352}, 849 (1991);
Nucl.\ Phys.\ B {\bf 360}, 641 (1991).

\bibitem{Kane:1992a}
C.~L.~Kane and M.~P.~A.~Fisher,
Phys. Rev. {\bf B46}, 15233 (1992).


\bibitem{Oshikawa:2005fh}
C.~Chamon, M.~Oshikawa, and I.~Affleck,
Phys.\ Rev.\ Lett.\  {\bf 91}, 206403 (2003).

\bibitem{callan90} 
C G. Callan, Jr. and L. Thorlacius, 
Nucl. Phys. B {\bf 329}, 117 (1990).

\bibitem{schmid} 
A. Schmid, 
Phys. Rev. Lett. {\bf 51}, 1506 (1983);
F. Guinea, V. Hakim, and A. Muramatsu,
Phys. Rev. Lett. {\bf 54}, 263 (1985).

\bibitem{fisher;1985}
M. P. A. Fisher and W. Zwerger, 
Phys. Rev. {\bf B32} 6190 (1985).

\bibitem{kane;1992b}
C. L. Kane and M. P. A. Fisher,
Phys. Rev. Lett. {\bf 68}, 1220 (1992).

\bibitem{furusaki}
A. Furusaki and N. Nagaosa,
Phys. Rev. {\bf B47}, 4631 (1993).

\bibitem{Callan:1994ub}
C.~G.~.~Callan, I.~R.~Klebanov, A.~W.~W.~Ludwig and J.~M.~Maldacena,
Nucl.\ Phys.\ B {\bf 422}, 417 (1994).

\bibitem{Polchinski:1994my}
J.~Polchinski and L.~Thorlacius,
Phys.\ Rev.\ D {\bf 50}, 622 (1994).

\bibitem{Sen:2002nu}
A.~Sen, 
JHEP {\bf 0204}, 048 (2002); 
See for a review on the rolling 
tachyon: A.~Sen, 
Int. J. Mod. Phys. {\bf A20}, 5513 (2005).

\bibitem{Tlee:06}
 T.~Lee,
JHEP {\bf 0611}, 056 (2006).

\bibitem{TLee;08}
T. Lee,
JHEP {\bf 02} 090 (2008).


\bibitem{Lee:2005ge}
T.~Lee and G.~W.~Semenoff,
JHEP {\bf 0505}, 072 (2005); 
M. Hasselfield, T. Lee, G. W. Semenoff, P. C. E. Stamp,
Ann. Phys. {\bf 321}, 2849 (2006).

\bibitem{Lee;mod}
T. Lee,
Int. J. Mod. Phys. {\bf A24}, 6141 (2009).

\bibitem{Lee;canonical}
T. Lee,
[arXiv:1507.08063].


\bibitem{Lee;2007}
S. Ji, J.-Y. Koo and T. Lee, 
J. Korean Phys. Soc. {\bf 50}, S54 (2007).

\bibitem{seib}
N. Seiberg and E. Witten,
JHEP 9909:{\bf 032}, (1999).

\bibitem{Lee;open}
T. Lee, 
Phys. Rev. {\bf D62}, 024022 (2000).

\bibitem{lee0105}  T. Lee, 
Phys. Rev. {\bf D64}, 106004 (2001).

\bibitem{para}
H. S. Green, Phys. Rev. {\bf 90}, 270 (1953); Y. Ohnuki and S. Kamefuchi, 
{\it Quantum Field Theory and Parastatistics} (University of Tokyo Press, 
Tokyo, springer, Berlin 1982).

\bibitem{quon} O. W. Greenberg, Phys. Rev. Lett. {\bf 64},
705 (1990); Phys. Rev. D{\bf 43}, 4111 (1991); M. Chaichian, R. Gonzalez Felipe and C. Montonen, J. Phys. A {\bf 26}, 4017 (1993);  S. Meljanac and A. Perica,
Mod. Phys. Lett. A {\bf 9}, 3293 (1994)
;V. Bardek, M. Doresic, and S. Meljanac, Phys. Rev. {\bf D} 49, 3059 (1994).



\end{thebibliography}
\end{document}